%% file: main.tex
\documentclass[journal]{vgtc}                
\graphicspath{{figures/}{pictures/}{images/}{./}} 

\usepackage{subfiles}
\usepackage{xspace}
\usepackage{multicol}
\usepackage{multirow}
\usepackage{booktabs}
\usepackage[svgnames]{xcolor}
\usepackage{makecell}

\def\measure{CLAMS\xspace}

\def\ambreducer{AmbReducer\xspace}

\onlineid{1026}

\vgtccategory{Research}
\vgtcpapertype{Theoretical \& Empirical}

\title{
\textit{\measure}: A Cluster Ambiguity Measure for Estimating \texorpdfstring{\\}{} Perceptual Variability in Visual Clustering}

\author{Hyeon Jeon*, Ghulam Jilani Quadri*, Hyunwook Lee, Paul Rosen, Danielle Albers Szafir, and Jinwook Seo}
\authorfooter{

\item Hyeon Jeon and Jinwook Seo are with Seoul National University. E-mail: hj@hcil.snu.ac.kr, jseo@snu.ac.kr
\item Ghulam Jilani Quadri and Danielle Albers Szafir are with the University of
North Carolina, Chapel Hill. E-mail: \{jiquad, danielle.szafir\}@cs.unc.edu
\item Hyunwook Lee is with UNIST. E-mail: gusdnr0916@unist.ac.kr
\item Paul Rosen is with the University of Utah. E-mail: prosen@sci.utah.edu
\item Hyeon Jeon and Ghulam Jilani Quadri equally contributed to the work.
}

\shortauthortitle{Biv \MakeLowercase{\textit{et al.}}: Global Illumination for Fun and Profit}

\abstract{
Visual clustering is a common perceptual task in scatterplots that supports diverse analytics tasks (e.g., cluster identification).
However, even with the same scatterplot, the ways of perceiving clusters (i.e., conducting visual clustering) can differ due to the differences among individuals and ambiguous cluster boundaries. 
Although such perceptual variability casts doubt on the reliability of data analysis based on visual clustering, we lack a systematic way to efficiently assess this variability.
In this research, we study perceptual variability in conducting visual clustering, which we call \textit{Cluster Ambiguity}.
To this end, we introduce \textit{\measure}, a data-driven visual quality measure for automatically predicting cluster ambiguity in monochrome scatterplots. 
We first conduct a qualitative study to identify key factors that affect the visual separation of clusters (e.g., proximity or size difference between clusters).
Based on study findings, we deploy a regression module that estimates the human-judged separability of two clusters.
Then, \measure predicts cluster ambiguity by analyzing the aggregated results of all pairwise separability between clusters that are generated by the module. 
\measure outperforms widely-used clustering techniques in predicting ground truth cluster ambiguity. Meanwhile, \measure exhibits performance on par with human annotators.
We conclude our work by presenting two applications for optimizing and benchmarking data mining techniques using \measure. The \textbf{interactive demo} of \measure is available at \texttt{\href{http://www.clusterambiguity.dev.s3-website.ap-northeast-2.amazonaws.com/}{clusterambiguity.dev}.}
} 


\definecolor{myblue}{RGB}{7, 115, 193}
\definecolor{myred}{RGB}{255, 10, 9}

\newcommand{\blue}[1]{{\color{myblue}#1}}
\newcommand{\red}[1]{{\color{myred}#1}}

\newcommand{\rev}[1]{{#1}}

\teaser{
  \centering
  \includegraphics[width=\linewidth]{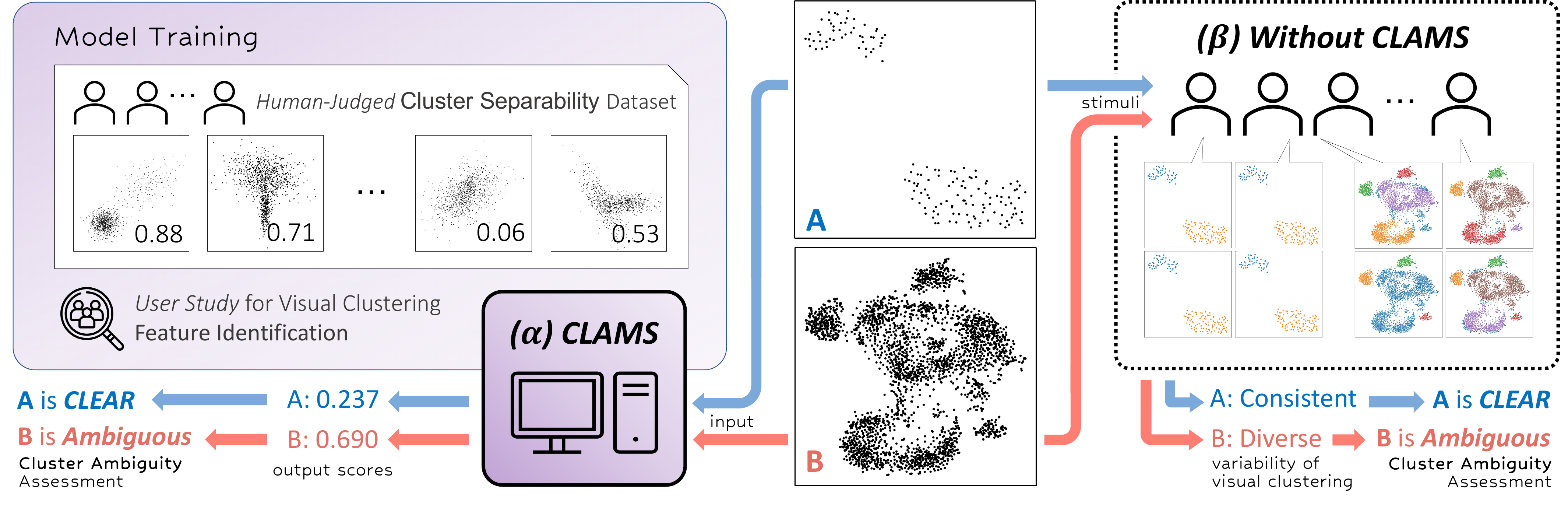}
  \caption{The comparison between the way of estimating the perceptual variability in conducting visual clustering, i.e., cluster ambiguity, of monochrome scatterplots. 
  Without \measure ($\beta$), we need to ask people to conduct visual clustering (colored scatterplots with $\beta$; color labels depict visual clustering results) and check whether the results are \blue{consistent (i.e., clear)} or \red{not (i.e., ambiguous)}. 
  This approach requires extensive time and human resources. 
  In contrast, \measure ($\alpha$), which is constructed over perceptual data and a feature engineering based on a user study, 
  automatically produces a score representing cluster ambiguity of an input scatterplot, 
  where \blue{low} and \red{high} scores correspond to \blue{clear} and \red{ambiguous} cluster structure. 
  }
  \label{fig:teaser}
}

\begin{document}


\firstsection{Introduction\label{sec:intro}}
\maketitle

\subfile{sections/01_introduction}
\subfile{sections/02_related_works}

\subfile{sections/03_measure}
\subfile{sections/04_evaluations}

\subfile{sections/05_applications}

\subfile{sections/06_conclusion}
\acknowledgments{
This work was supported by NAVER Corporation (Cloud Data Box), by the National Research Foundation of Korea (NRF) grant funded by the Korean government (MSIT) (No. 2023R1A2C200520911), and by the National Science Foundation under grant No. 2127309 to the Computing Research Association for the CIFellows project, NSF IIS-2046725, NSF IIS-1764089, NSF IIS-2320920, NSF IIS-2233316, and NSF III-2316496.
The authors wish to thank Sungbok Shin, Yun-Hsin Kuo, Seokhyeon Park and SNU-HCIL / UNC-VisuaLab members for their valuable feedback. We would also like to appreciate Micha\"el Aupetit for providing ClustMe dataset.}

\bibliographystyle{abbrv-doi}

\bibliography{ref}
\end{document}

%% file: sections/01_introduction.tex
Clustering is a key analytic task in scatterplots~\cite{quadri21tvcg, xia22tvcg, xia21cgna}.
It occurs when we infer structure in data by identifying groups (i.e., clusters) based on the pairwise proximity between data points~\cite{amar2005low}.
Visual clustering occurs when people infer these groups visually, such as finding neighborhoods of points in a scatterplot.
It is among the most common perceptual tasks people conduct with scatterplots~\cite{sarikaya2018scatterplots}.
Diverse domains, such as bioinformatics~\cite{shi10cbm, shannon03phar} and machine learning~\cite{kahng18tvcg}, leverage visual clustering for data analysis. These applications include the identification of ground truth clusters for benchmarking data mining techniques, such as automatic clustering~\cite{aupetit19vis, xia21cgna} and dimensionality reduction techniques~\cite{etemadpour2014perception, xia22tvcg}.
%

While visual clustering supports a range of applications, 
\textit{cluster ambiguity}---the intrinsic perceptual variability in visual clustering due to unclear cluster boundaries---can introduce uncertainty or even errors in applications relying on visual clustering. 
%
For example, 
cluster ambiguity can make data analysis unreliable. If a scatterplot is highly ambiguous,
it is easy to make multiple conclusions about cluster structure within the data. 
%
%
Ambiguity also reduces the reliability of establishing ground truth clusters based on human perception for benchmarking data mining techniques: if each person perceives cluster structure differently, we cannot know which structure is ``correct.''

However, most studies and models of visual clustering focus on how people perceive clusters in general~\cite{abbas19cgf, xia21cgna, wertheimer22pf}. For example, Gestalt principles~\cite{wertheimer22pf} explain how people ``commonly'' or ``generally'' group visual objects within a complex scene~\cite{pinna2010new}. 
Studies examining average cluster perception support visualization designers in building effective systems 
that most people can use~\cite{lu20tvcg, wang19tvcg, nonato19tvcg}. However, they do not offer solutions to deal with the challenges in data analysis and benchmarking imparted by cluster ambiguity. Designers thus need a more systematic way to assess cluster ambiguity.

A reliable method to evaluate the ambiguity of a scatterplot is directly measuring perceptual variability via human experiments~\cite{hartwig2023clusternet}.
However, this process is costly and not scalable.
As an alternative, we can use clustering techniques to mimic human perception \cite{aupetit2016sepme, sedlmair2015data} (\autoref{sec:vqmcp}). 
Assuming that a clustering technique with a specific hyperparameter setting represents a human’s perception, this method mimics human variability by running the technique under various hyperparameter settings.
 This approach is scalable but at the cost of 
accuracy due to the lack of 
human input (\autoref{sec:mainstudy}).


This research presents a scalable and accurate method to evaluate cluster ambiguity through a visual quality measure (VQM) called \textit{\measure}. 
We design \measure based on a dataset gathered from human input about the separability of clusters~\cite{abbas19cgf}. 
We construct \measure in two steps: first, we conduct a user study to investigate important factors that influence visual clustering. 
Second, based on the study findings, we train a regression module estimating how human subjects separate clusters.
Given a scatterplot, \measure computes the separability of every pair of identified clusters using the regression module. 
The measure then aggregates the computed pairwise separabilities of clusters to predict how ambiguously the clusters are portrayed by human subjects. 


Our quantitative experiments show that \measure is more accurate than existing models in predicting cluster ambiguity.
First, an ablation study verifies the accuracy of the regression module, validating our user-study-driven approach in constructing the module.
Moreover, we find that the ranking of scatterplots set by \measure has a strong correlation with the ground truth ranking constructed from 20 participants in our study.
Furthermore, \measure outperforms an average human annotator in estimating cluster ambiguity.
We also present two applications of \measure in optimizing and benchmarking data mining algorithms. 
First, we propose \ambreducer, an optimization system that reduces the ambiguity of dimensionality reduction embeddings while maintaining accuracy.
\ambreducer helps analysts effectively interpret high-dimensional data by informing cluster ambiguity. 
Second, we show how our measure can help select reliable benchmark datasets for  
comparing different clustering techniques. 
Findings from our experiments and applications open up discussions on leveraging perceptual variability in 
visualization research. 

%% file: sections/02_related_works.tex
\section{Background and Related Work}
We survey past work about visual clustering in scatterplots,  visual quality measures, and perceptual variability. 
These works collectively illustrate the necessity of \measure.

\subsection{Cluster Perception in Scatterplots}

\label{sec:cps}

Given the 
broad applications for visual clustering,
several previous works have concentrated on understanding and modeling cluster perception.
For example, eye-tracking 
was used to detect what aspects of a dataset people use for
cluster identification, highlighting the role of Gestalt principles, especially proximity and closure~\cite{etemadpour2014eye}. 
ScatterNet captures perceptual similarities between scatterplots to emulate human clustering decisions~\cite{ma2018scatternet}. 
Scagnostics identify scatterplot patterns, including cluster structure~\cite{dang2014transforming}, but 
cannot reliably reproduce human perception~\cite{pandey2016towards}. 
In ClustMe~\cite{abbas19cgf},
Abbas et al. 
\rev{computationally} modeled \rev{human perception in judging the} complexity of the cluster structure in scatterplots, which affects by the number of clusters and the nontriviality of patterns.
Quadri \& Rosen studied various factors that influence the perception of clusters. Relying on these factors, they built explainable models of how humans perceive cluster separation based on merge tree data structures from topological data analysis~\cite{quadri21tvcg}.

These studies aim to 
characterize cluster perception processes, 
the factors that influence cluster perception, and the relation between these factors and scatterplot design. However, 
visual clustering in scatterplots 
lacks any ground truth; we cannot always categorically determine which group of points forms a cluster. 
Underlying data characteristics lead to ambiguity in cluster structure (i.e., cluster ambiguity), causing intrinsic variability in 
the clusters people detect.
We thus develop a VQM that automatically estimates cluster ambiguity
for a wide range of cluster patterns. We then explore and rank the cluster ambiguity of scatterplots.

\subsection{Visual Quality Measures on Cluster Perception}

\label{sec:vqmcp}


The visualization community has proposed and developed various \textit{Visual Quality Measures} (VQMs)~\cite{bertini2011quality, behrisch2018quality} 
that metricize the specific characteristics of cluster patterns in scatterplots.
These metrics enable analysts to rapidly evaluate their scatterplots in terms of supporting general pattern exploration \cite{brehmer2013multi} or a specific perceptual task \cite{amar2005low}, and can also be used to optimize scatterplots\cite{quadri22tvcg}. 
%

A na{\"i}ve approach in designing VQMs is to reuse existing algorithms (e.g., clustering techniques), 
where the performance of the algorithm is evaluated against the 
results of human experiments.
Aupetit et al.\ investigated how well clustering techniques can mimic (i.e., model) visual clustering by testing the extent to which 1,400 variants of clustering techniques can reproduce the human separability judgments~\cite{aupetit19vis}. 
Sedlmair \& Aupetit proposed a framework for evaluating class separability measures based on human-judged class separability data and 
used the framework to test 15 class separation measures~\cite{sedlmair2015data}.
The framework was later extended to test a large set of class separability measures, which were constructed as combinations of 17 neighborhood graph functions and 14 class purity functions~\cite{aupetit2016sepme}. However, reusing existing algorithms does not directly reflect human perception, thus often performing poorly in imitating human perception~\cite{abbas19cgf} (\autoref{sec:mainstudy}).

VQMs can be instead designed to target a specific pattern or a perceptual task. For example, 
Quadri \& Rosen~\cite{quadri21tvcg}
developed a threshold plot based on their modeling of cluster perception.
Threshold plots tell the extent to which clusters are salient in an input scatterplot.
Designers can use these plots to 
enhance the effectiveness and efficiency of a scatterplot for visual clustering~\cite{quadri22tvcg}. 
ClustMe~\cite{abbas19cgf} also produced a VQM
simulating human perception in judging the complexity of cluster patterns. Specifically, ClustMe models human perception based on 
experimental data on perceived cluster separability judged by multiple subjects. 
We 
apply this modeling approach and the perceptual data from ClustMe to guide our model.
However, ClustMe regards the judgments of multiple subjects as a binary decision (i.e., separated or not separated), estimating a \textit{general} complexity perceived. In contrast,
\measure summarizes the judgments made by subjects as a continuous probabilistic function for the sake of estimating the \textit{variability} of the judgments  (i.e., cluster ambiguity).
\rev{Estimating variability helps us understand the distribution of human visual percepts, enabling related applications to more accurately reflect perception.}

\subsection{Perceptual Variability}
Due to perceptual variability among individuals, the perception of clusters can differ even with the same scatterplot.  Perceptual variability refers to individuals’ ``traits or stable tendencies to respond to certain stimuli or situations in predictable ways''~\cite{dillon1996user}. Prior work on perceptual variability demonstrated that people exhibit observable differences in task-solving and behavioral patterns~\cite{woolhouse2000personality}. 
Psychology and vision research 
have shown that 
people exhibit substantial variability 
on tasks such as pattern identification, judgments, or adjustments~\cite{hoskin2019sensitivity, huang2010distortions, ons2011subjectively}. 
Given the significance of perceptual variability in information visualization~\cite{davis2022risks,ziemkiewicz2009preconceptions}, 
 recent research explores when, where, and how perceptual variability influences visualization use (see Liu et al.~\cite{liu2020survey} for a survey).

\begin{figure*}
    \centering
    \includegraphics[width=\linewidth]{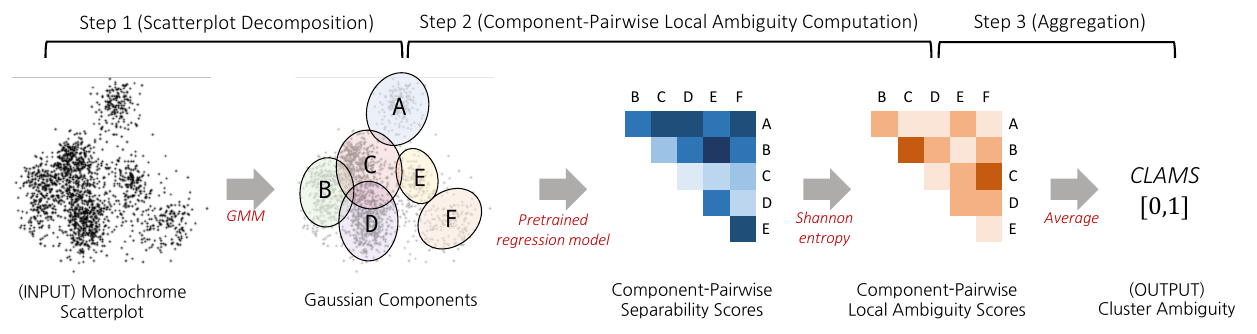}
    \vspace{-6mm}
    \caption{The 
    \measure pipeline (\autoref{sec:pipeline}). (Step 1) Given a scatterplot, we first apply a Gaussian Mixture Model (GMM) to abstract 
    a scatterplot into a set of Gaussian components. (Step 2) We estimate component-pairwise separability scores by applying a predefined regression module (\autoref{sec:regmodel}) and convert the scores into local ambiguity scores by applying the binary Shannon entropy function. (Step 3) Finally, we obtain the cluster ambiguity score of the input scatterplot by averaging the local ambiguity scores.
    \vspace{-4mm}
    }
    \label{fig:pipeline}
\end{figure*}

Perceptual variability can influence the reliability of any generalized model of perception~\cite{zaman21pbr}. 
The Axiom of Perceptual Variability
suggests that the success of such generalized perceptual theories depends on their ability to account for variance in perceptual representations~\cite{ashby1993perceptual}. 
Moreover, the perceptual variability of a certain task may degrade the credibility of the applications relying on visual clustering (\autoref{sec:intro}).
Visual clustering is an ill-posed problem, where 
there is no "ground truth" for clusters 
 (i.e., it is not always possible to determine a ``correct'' clustering). 
Generalized models and applications of visual clustering are thus likely to be more vulnerable to perceptual variability. We believe that our modeling of perceptual variability in visual clustering (i.e., cluster ambiguity) not only enhances our understanding of cluster perception but also resolves \rev{the} vulnerabilities \rev{of such models and applications}.

%% file: sections/03_measure.tex
\section{\measure}


We introduce \measure, a VQM for estimating the cluster ambiguity of a monochrome scatterplot.
Trained over perceptual experiment data, our approach \rev{acts} as a proxy for human perception.
Moreover, by decomposing an input scatterplot with a Gaussian mixture model (GMM) and measuring ambiguity in a component-pairwise manner, \measure can deal with a wide range of cluster patterns while maintaining scalability. 

\subsection{Design Considerations}

\label{sec:decon}

Based on a thorough examination of the related work on cluster perception (\autoref{sec:cps}) and VQMs (\autoref{sec:vqmcp}), we set three design considerations in estimating cluster ambiguity.

\setlist{topsep=0.1em}
\begin{itemize}[noitemsep] 
    \item[\textbf{(C1)}] \textbf{Provide a proxy for human perception:} To assess the variability of people's perception, the measure should accurately reflect how people identify and analyze clusters in practice \cite{quadri21tvcg, abbas19cgf, aupetit19vis}. 
    Refer to Step 2 in \autoref{sec:pipeline} for how we achieved this consideration.
    \item[\textbf{(C2)}] \textbf{Work with a wide range of scatterplot patterns:} Scatterplots
    have a wide range of cluster patterns; they can have clusters with diverse characteristics (e.g., shape, density, size)~\cite{jeon22tvcg} and varying numbers of clusters~\cite{quadri21tvcg}. 
    \measure should 
    properly estimate the ambiguity of an arbitrary scatterplot and should properly estimate cluster ambiguity across a diverse array of patterns
    \cite{quadri21tvcg, abbas19cgf}. 
    Check Step 1 in \autoref{sec:pipeline} to see how our measure design regards this consideration.
    \item[\textbf{(C3)}] \textbf{Be scalable:} To be readily used in practice (\autoref{sec:appl}), 
    \measure should scale to large numbers of data points and complex patterns \cite{abbas19cgf}. 
    Throughout the entire pipeline (Steps 1, 2, and 3), we maintain scalability as a key consideration.
\end{itemize}

\subsection{\measure Pipeline}

\label{sec:pipeline}

We draw on previous approaches (e.g., VQM~\cite{abbas19cgf}, clustering~\cite{halkidi01icdm, rousseeuw87jcam}, dimensionality reduction~\cite{venna06nn, lee2007springer}) to design our measure to (Step 1) decompose a given scatterplot into smaller components, (Step 2) compute component-pairwise local ambiguity scores, and (Step 3) estimate the global ambiguity score by aggregating the local scores (see \autoref{fig:pipeline}). As the diversity of local components is 
inherently much lower than 
that of the full scatterplot, this decomposition
helps \measure to readily work with a wide range of scatterplot patterns (C2). 

\subsubsection*{(Step 1) Scatterplot Decomposition}

\measure starts by decomposing an input scatterplot into smaller components. 
We use the Gaussian Mixture Model (GMM), which decomposes a given dataset as a mixture of multidimensional Gaussian distributions. 
We select GMM due to several advantages. 
First, GMM does not require hyperparameter selection~\cite{abbas19cgf}. The number of Gaussian distributions (components) can be automatically determined based on fixed statistics, such as Bayesian information criteria (BIC)~\cite{schwarz78ans}. Moreover, as each component is represented as a Gaussian distribution, complex patterns can be abstracted into a concise statistical summary (e.g., mean, covariance matrix; C2).
Note that \rev{GMM decomposition has been shown to be applicable to visual identification problems \cite{abbas19cgf, abbas21arxiv}, accurately representing a wide range of smooth cluster patterns.}
Finally, the complexity of GMM for 2D scatterplots is $O(NK)$, where $N$ and $K$ denote the number of points and components, respectively, ensuring that the technique is highly scalable (C3).

In contrast, a conventional approach for scatterplot decomposition, which involves using clustering techniques (e.g., $K$-Means~\cite{likas03pr}, HDBSCAN~\cite{mcinnes17icdmw}), falls short in enabling \measure to satisfy our target considerations.
First, clustering results can vary significantly due to the sensitivity of the outcomes to changes in hyperparameters. We can find an optimal hyperparameter setting using clustering validation measures~\cite{xiong2018clustering}, but the optimal setting may depend on the selection criteria~\cite{liu10icdm}. Clustering techniques also provide a partition of data points, meaning they neither abstract nor simplify clustering patterns. Lastly, clustering techniques that are widely known to be able to capture complex patterns (e.g., density-based clustering \cite{mcinnes17icdmw}, density-peak clustering \cite{liu18is}) often suffer from scalability issues.

Note that while applying GMM to the input scatterplot, we determine the optimal number of Gaussian components based on BIC scores and the elbow rule. The elbow is found using the Kneedle~\cite{satopaa11icdcsw} algorithm.

\subsubsection*{(Step 2) Component-Pairwise Local Ambiguity Computation}

We 
predict local ambiguity for every pair of Gaussian components so that \measure can consider every possible interaction between 
components. 
This process is as follows: 

\noindent
\textbf{Dataset:}
We use the human-judged cluster separability dataset $\mathbf{X}$ from the ClustMe study~\cite{abbas19cgf} (C1). $\mathbf{X}$ contains 1,000 scatterplots $\{X_1, X_2, \cdots, X_{1000}\}$, where each scatterplot consists of a pair of Gaussian components with diverse statistics (e.g., mean, covariance). 
The separability scores $\mathbf{S} = \{S_1, S_2, \cdots, S_{1000}\}$ are computed by aggregating the judgments of 34 participants on each scatterplot gathered by the ClustMe study~\cite{abbas19cgf}. These scores reflect the judgments of the participants 
 regarding whether they 
saw one or multiple clusters in each scatterplot. 
$S_i$ represents the proportion of participants who perceived more than one cluster in $X_i$. 

\noindent
\textbf{Deriving Cluster Ambiguity from Separation Score:}
We compute the cluster ambiguity of each Gaussian components pair with separation score $S$ 
as: $A(S) = - S \log S - (1-S) \log (1-S)$. Theoretically, $A$ is defined as the Shannon entropy of a binomial distribution representing the probabilistic event of counting the number of clusters. We use entropy as it is an efficient and mathematically grounded function representing the level of ``uncertainty'' of the associated event~\cite{massen88aps}
(C3). 
$A$ is minimized when $S$ is 0 or 1 (i.e., every participant 
gave the same answer, meaning there was no variability in visual clustering), and is maximized when $S$ is 0.5 (i.e., half of the participants saw a single cluster and the other half saw more than one cluster, maximizing variability). 
As the scatterplots and original separability scores have proven to be effective in modeling cluster perception~\cite{abbas19cgf}, we can also expect the ambiguity scores 
to reliably represent human perception. Note that our evaluation validates this expectation (\autoref{sec:mainstudy}; C1). 

\noindent
\textbf{Predicting Ambiguity:}
To predict local ambiguity for each Gaussian component, we train a regression module $f$ estimating the separability score $f(X)$ of an arbitrary pair of Gaussian components $X$, using $\mathbf{X}$ and $\mathbf{S}$. For an arbitrary pair of Gaussian components $X$, the local ambiguity score of a pair is computed as $A(f(X))$. 
Although 
training the regression module requires heavy computation (\autoref{sec:regmodel}), 
inferring local ambiguity scores can be done in constant time regardless of the number of Gaussian components, making Step 2 scalable (C3). 
Please refer to \autoref{sec:regmodel} for a detailed explanation of how we designed and implemented the model.

\subsubsection*{(Step 3) Aggregating Local Ambiguity}

Finally, we aggregate the component-pairwise ambiguity scores by averaging the scores.
Note that the final score can be interpreted as the joint entropy of a set of events corresponding to each local ambiguity score with an assumption that the events are mutually independent.

\subsection{Computational Complexity}

As the time complexity of GMM on a 2D scatterplot is $O(NK)$, the time complexity of Step 1 is $O(NK_{max}^2)$, where $K_{max}$ is the maximum number of components we consider in finding the optimal number of components and $N$ is the number of points. The complexity for both Steps 2 and 3 is $O(K_{opt}^2)$, where $K_{opt}$ denotes the optimal number of components found in Step 1. Thus, the computational complexity of \measure is $O(NK_{max}^2 + K_{opt}^2)$. As $N \gg K_{max}$ and $N \gg K_{opt}$, the effect of $K_{max}$ and $K_{opt}$ is negligible, making \measure a scalable measure that is only linear to the number of points. Refer to Appendix B for the quantitative experiment demonstrating the scalability of \measure.

\section{Estimating Human-Judged Cluster Separability}

\label{sec:regmodel}

We constructed a regression module to predict the separability of a given Gaussian components pair. 
We conducted a user study exploring the factors affecting visual clustering to ground our model in human reasoning processes. We trained the model to estimate ambiguity from the features extracted from each pair, where the features were engineered based on the study findings.

\definecolor{mypurple}{RGB}{123, 64, 167}
\definecolor{myorange}{RGB}{255, 118, 38}

\newcommand{\purple}[1]{{\color{mypurple}#1}}
\newcommand{\orange}[1]{{\color{myorange}#1}}

{\renewcommand{\arraystretch}{1.5}

\begin{table*}[t]
    \centering
    \resizebox{\linewidth}{!}{
    \begin{tabular}{c|lllc}
     \toprule
     \multirow{7}{*}{\includegraphics[height=1.45in]{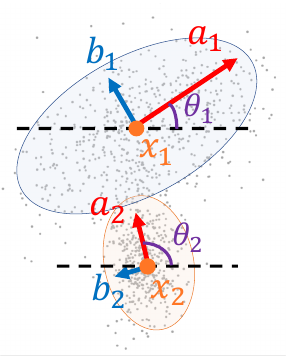}} 
         & \textbf{\textit{Factors}} &   \textbf{Features}       & \textbf{Formula}  & \textbf{Used} \\
    \cline{2-5}
        & \textit{Proximity} & Distance between centers (DC) & $\parallel x_1 - x_2 \parallel$ & \small{$\bigcirc$}\\ 
        & \textit{Clarity of a gap} & Distance-size ratio (DSR) &  $\parallel x_1 - x_2 \parallel / (\sqrt{\parallel a_1\parallel ^2 + \parallel b_1 \parallel ^2} +\sqrt{\parallel a_2\parallel ^2 + \parallel b_2 \parallel ^2} )$ & \small{$\bigcirc$}\\ 
        & \textit{Density Difference} & Density difference (DD) & $ \mid (n_1 / (2 \cdot \parallel a_1\parallel \cdot \parallel b_1\parallel)) - (n_2 / (2 \cdot \parallel a_2\parallel \cdot \parallel b_2\parallel)) \mid$ & $\times$ \\
        & \textit{Size Difference} & Size difference (SD) & $\mid \sqrt{\parallel a_1\parallel ^2 + \parallel b_1 \parallel ^2} -\sqrt{\parallel a_2\parallel ^2 + \parallel b_2 \parallel ^2} \mid $ & \small{$\bigcirc$}\\
        & \textit{Ellipticity Difference} & Ellipticity difference (ED) & $\mid (\parallel a_1\parallel / \parallel b_1\parallel - \parallel a_2\parallel / \parallel b_2\parallel )\mid$&\small{$\bigcirc$} \\
        & \textit{Direction}&  Angle between components (AC) & $ \min(\Delta_\theta, 2\pi - \Delta_\theta)$ where $\Delta_\theta = \mid \theta_1 - \theta_2 \mid$ & \small{$\bigcirc$}\\
         \bottomrule
        \addlinespace[0.05cm]
        \multicolumn{5}{l}{
        \footnotesize
            $a_1, a_2$ \& $b_1, b_2$: Standard deviation along the major \& minor axis of Gaussian components / $x_1, x_2$: the center of Gaussian components
            
         \vspace{-1.78mm}
        } \\ 
        \multicolumn{5}{l}{
        \footnotesize
        $\theta_1, \theta_2$: 
Angle between the major axis of Gaussian components and the horizontal line / $n_1, n_2$: number of points in Gaussian components
        }
    \end{tabular}
    }
    \vspace{-2mm}
    \caption{The 
    visual clustering factors revealed by our preliminary user study (\autoref{sec:preexp}; first column) and the corresponding features designed for the regression module predicting the human-judged separability of Gaussian components (\autoref{sec:feateng}; second column). The third column depicts how we compute each feature, and the fourth column represents whether or not the feature is used in the final regression module (\autoref{sec:regmodeleval}). Empowered by the designed features, the module succeeds in precisely estimating the separability scores.  \vspace{-4mm}}
    \label{tab:feature_eng}
\end{table*}

}

\subsection{Factor Exploration Study}
\label{sec:preexp}

To design features that relate to the human-judged cluster separability,
we first explored 
factors that affect visual clustering via a qualitative experiment. In the study, 
participants completed a series of visual clustering tasks and reported on 
factors that affected task results. 

\subsubsection{Study Design}

\noindent
\textbf{Procedure and Tasks}: 
The experiment began with informed consent and a basic demographic survey. Participants completed the rest of the study in three phases: (1) identifying clusters in a single scatterplot, (2) performing pairwise comparisons between scatterplots, and (3) participating in an interview to elicit core factors influencing visual clustering. 
In the first phase, we randomly sampled 12 scatterplots from $\mathbf{X}$ and showed them in sequence. For each scatterplot, we first asked participants to select whether there existed one or more than one cluster, following the original ClustMe~\cite{abbas19cgf} study. We additionally asked participants about their confidence in 
their response using a Likert scale
and asked them to describe the reasoning process
both for the number of clusters and their confidence. 

In the second phase, we randomly sampled 12 pairs of scatterplots (24 in total) from $\mathbf{X}$. 
For each trial, we asked participants to report which scatterplot in a given pair was more separated. As in the first phase, we asked for the participants' confidence and had them report the reasoning behind their choices and confidence. 
After the two phases of the experiment, we conducted a semi-structured post hoc interview, asking the participants about the salient factors they felt were affected by cluster separation.
All participants finished the experiment within one hour.

\noindent
\textbf{Participants:}
We recruited 10 participants from a local university (six males and four females, aged 19--28 [$23.5\pm 2.6$]). Six of the participants were undergraduates, three were graduate students, and one had just completed their Bachelor's degree.
We selected only participants who had experience in data analysis using scatterplots to better ground our results in real-world data analysis and to ensure that they understood the concept of clusters.
\rev{Three participants reported that they are at a novice level in data analysis, which meant that they did not regularly conduct analyses but have some experience. Five and two participants reported themselves as at intermediate and expert levels, respectively. Half of the participants reported being at the novice level of data analysis using a scatterplot and the other half reported themselves as intermediate.}
\rev{We also confirmed that participants had not read three papers \cite{abbas19cgf, abbas21arxiv, aupetit19vis} that incorporate ClustMe data (i.e., $\mathbf{X}$). }
Participants were compensated with the equivalent of \$20.

\noindent
\textbf{Apparatus:}
The experiment was conducted over Zoom, and the sessions were recorded. We developed a website in which participants could see scatterplots and make their selection (number of clusters and confidence) with a mouse click. 
\rev{
We fixed the stimuli size to 700px $\times$ 700px and constrained participants to use a laptop or desktop screen to minimize the impact of the display on study results.
}
They were asked to access the website and share their screens so that the experimenter could monitor and guide the experiment. 

\subsubsection{Results}
\label{sec:preexpresults}

We analyzed the responses from the main experiment and the post hoc interview using axial coding done by two authors, one of whom works in machine learning while the other's primary expertise is in visualization.
The coders individually developed a separate codebook in the first stage. Two codebooks were then merged, 
resulting in 13 total extracted factors. Finally, based on two stages of discussions, the coders agreed to categorize the codes into six main factors
(see \autoref{tab:feature_eng})---proximity, clarity of gap, density difference, size difference, ellipticity difference, and direction.

The most important factor was the \textit{proximity} between the clusters, 
which was mentioned by seven out of ten participants in the interviews. 
The perception of proximity was mainly affected by the \textit{clarity of a gap} between clusters. Participants reported that if the gap between clusters was bigger and 
less dense than the clusters, the clusters were perceived to be more distinct. Five out of ten participants explicitly mentioned the gap between clusters as an important factor in the interviews. 

Participants tended to perceive more than one cluster when two Gaussian components had noticeably \textit{different densities}, even if there was no gap between them or if they overlapped (i.e., when they had high proximity). However, two participants explicitly noted that if the density of a point group was too low, the points within the group 
were perceived as outliers rather than a cluster. We found that the \textit{size difference} between clusters similarly affects visual clustering: 
if a cluster was too small, participants often wanted to interpret the cluster as a set of outliers. In the 
interviews, 30\% (three out of 10) of participants noted that the differences in density were salient features that affected their choices, whereas 
one noted that differences in cluster size played a role in their decisions.

\textit{Ellipticity difference} between clusters 
also affected perceived separability.
Participants mentioned that if the ellipticity of a cluster was high, it 
could be fit by linear regression where the regression line accorded with the major axis of the ellipse, thus would be more likely to be perceived as a single cluster. 
We also found that if the two clusters had high ellipticity, their \textit{direction} (i.e., the direction of the first principal axis) also played a crucial role in their separation. 
Participants noted that two clusters with high ellipticity and different directions were likely to fit into two independent regression lines, thus more likely to be perceived as two independent clusters. 
In the 
interviews, five out of ten participants mentioned ellipticity and direction as salient factors, while four additionally mentioned linear regression or correlation lines. 

\subsection{Feature Engineering} 

\label{sec:feateng}

We designed features representing the characteristic of a pair of Gaussian components based on the six factors 
from the study (\autoref{sec:preexp}). 
To maintain the simplicity and explainability of the model, we aimed to keep the features as simple and few as possible.
We thus designed the features to be (1) be bijective to the factors and (2) directly computed from the statistical summary of the Gaussian components pair (see \autoref{tab:feature_eng}). 
The features are as follows: 

\noindent
\textbf{Distance between Centers (DC):}
For 
\textit{proximity}, we selected the distance between the centers of Gaussian components. We used the distance between centers as it is the sole metric representing proximity that can be directly derived from the population statistics (i.e., mean) and is also widely used for measuring the proximity between clusters~\cite{liu10icdm}. 

\noindent
\textbf{Distance-Size Ratio (DSR):}
The gap between two Gaussian components generally becomes bigger if (1) the distance between the center of two Gaussian components increases or (2) the size of the components becomes smaller. 
We used this property to construct a feature corresponding to the \textit{clarity of the gap} as the ratio of the distance between centers of Gaussian components over the sum of the size of two components. Note that we defined the size of a component as its standard deviation---the root sum square of the standard deviations along the first and second principal axes.

\noindent
\textbf{Density Difference (DD):}
The \textit{density difference} factor can be directly represented
by computing the density of a component as the ratio of the number of points over the area covered by the component. 
We defined the area of a component as 
that of an ellipse where the major and minor axes' length is identical to the standard deviation along the first and second principal axes, respectively. 

\noindent
\textbf{Size Difference (SD):}
\textit{Size difference} 
can be also directly represented as a feature. We defined the size of the component as its standard deviation, as we do for the distance-size ratio feature.

\noindent
\textbf{Ellipticity Difference (ED):}
We directly used \textit{ellipticity difference} 
as a feature, 
defining the ellipticity of a component as the one of an ellipse having the standard deviation along the first and second principal axes as major and minor axes' lengths, respectively.

\noindent
\textbf{Angle between Components (AC):}
Participants reported \textit{direction} as salient only if they think two components are heading toward different directions. 
We therefore define direction as the angle between components, which can be represented as the angle between the first principal axes of two components. 

\subsection{Modeling Training} 

We trained a regression module for estimating human-judged separability of Gaussian components pair using $\mathbf{X}$ and $\mathbf{S}$. 
\rev{We used a regression module as we aimed to predict continuous scores.}
For a given pair of Gaussian components, the model first extracts the features as defined above and then predicts a separability score from the features. 
We implemented our module using AutoML~\cite{he21kbs} supported by the \verb|auto-sklearn|~\cite{feurer15nips} library. 
We report the performance of our model and the significance of the extracted features in \autoref{sec:regmodeleval}.

%% file: sections/04_evaluations.tex
\begin{figure*}
    \centering
    \includegraphics[width=\linewidth]{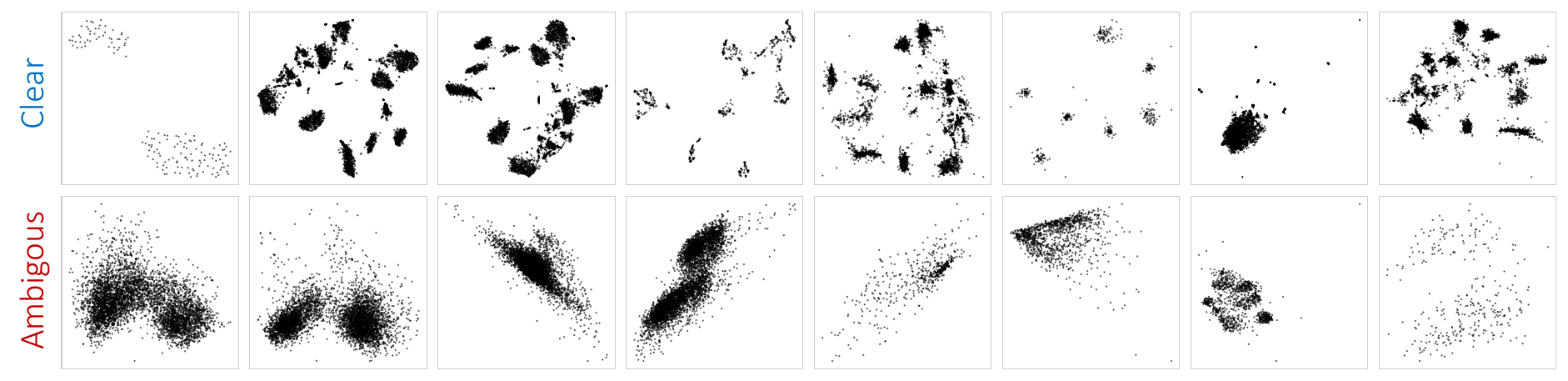}
    \vspace{-5mm}
    \caption{The top eight \blue{clear} scatterplots (low cluster ambiguity; first row) and the top eight \red{ambiguous} scatterplots (high cluster ambiguity; second row) based on \measure score, which are picked within 60 scatterplots we used in the accuracy evaluation (\autoref{sec:mainstudy}). \vspace{-4mm}}
    \label{fig:topbottom}
\end{figure*}

\section{Quantitative Evaluation}

We conducted two evaluations to demonstrate the validity and effectiveness of \measure. We first  examined the performance of our regression module and the significance of its features through an ablation study (\autoref{sec:regmodeleval}). We then evaluated the accuracy (\autoref{sec:mainstudy}) of \measure by comparing it against both computational methods for estimating cluster ambiguity and human annotators. 

\subsection{Ablation Study for the Regression Module}

\label{sec:regmodeleval}

We report the results of the ablation study examining (1) the performance of our regression module (\autoref{sec:regmodel}) and (2) the significance of the features we designed (\autoref{tab:feature_eng}; \autoref{sec:feateng}).

\subsubsection{Study Design}

\textbf{Objectives and Design:}
The study aimed to achieve two objectives: (1) to investigate the accuracy of our regression module in estimating human-judged separability and (2) to analyze 
how much each feature contributed to the model performance (i.e., the significance of features). 
For the first objective, we evaluated the performance of our model as it is (i.e., used all features we discovered). To accomplish the second objective, we switched off each feature individually and examined the extent to which the performance deteriorated. Additionally, to take into account the interplay between features, we repeated the process by disabling feature pairs.

\noindent \textbf{Measurement:}
We conducted a five-fold cross-validation to examine the performance of the model while using $\mathbf{X}$ and $\mathbf{S}$ as input and target, respectively. 
We used $R^2$ for the performance metric as it is interpretable~\cite{cameron97joe} and, moreover, unbiased if the number of points is fixed (as in our case).


\subsubsection{Results and Discussions}

{\renewcommand{\arraystretch}{1.2}

\begin{table}[t]
    \centering
    \begin{subtable}[]{\linewidth}
        \resizebox{\linewidth}{!}{
        \begin{tabular}{rccccccc}
         \toprule
            & & \multicolumn{6}{c}{Excluded feature} \\
            \cmidrule{3-8}
            & \textit{orig.} & DC & DSR & DD & SD & ED & AC \\
            \midrule
            $R^2$ & .9106 & .9075  & .8574 & .9205 & .8790 & .8758 & .9019  \\ 
            Change & 0\% & -0.34\% &  -5.84\% & +1.08\% & -3.47\% & -3.82\% & -0.95\% \\  
         \bottomrule
        \end{tabular}
        }
        \caption{Model performance with entire (\textit{orig.}) features or without a single feature\vspace{1.5mm}}
    \end{subtable}
    \begin{subtable}[]{\linewidth}
        \resizebox{\linewidth}{!}{
        \begin{tabular}{rrcccccc}
         \toprule
            & & \multicolumn{6}{c}{Excluded feature} \\
            \cmidrule{3-8}
                                     &      & DC      & DSR     & DD      & SD      & ED      & AC     \\
            \midrule
            \multirow{6}{*}{\makecell{Excluded \\ Feature}} & DC   &         & .5817   & .8751   & .8917    & .8495   & .8333  \\
                                     & DSR  & -36.1\% &         & .8521   & .8985    & .8560   & .8902  \\
                                     & DD   & -3.89\% & -6.42\% &         & .8925    & .8790   & .8721  \\
                                     & SD   & -2.08\% & -1.33\% & -1.99\% &          & .8750   & .9041 \\
                                     & ED   & -6.71\% & -6.00\% & -3.47\% & -3.91\%  &         & .8888  \\
                                     & AC   & -8.49\% & -2.24\% & -4.23\% & -0.71\%  & -2.40\% & \\
         \bottomrule
        \end{tabular}
        }
        \caption{Model performance without a pair of features (in corresponding row and column)}
        
    \end{subtable}

    \caption{The result of an ablation study (\autoref{sec:regmodeleval}) analyzing the performance of a regression module estimating human-judged separability scores. We seek how the performance of the module varies as we remove a single (a) or a pair (b) of features, reporting the $R^2$ score (first row in (a), upper right half in (b)), and the change rate compared to the model using entire features (second row in (a), lower left half in (b)). }
    \label{tab:ablation}
\end{table}
}

\textbf{Model Performance:}
As in \autoref{tab:ablation} (a), the model achieved $R^2$ score over 0.9 with all features (which we mark as \textit{orig.}). 
This result validates the effectiveness of the features and also the reliability of our study-driven approach to designing features.

\noindent \textbf{Significance of the Features:}
According to \autoref{tab:ablation} (a), DSR caused the biggest degradation of the performance when removed. ED was next, and SD was third. Meanwhile, removing DC, DD, and AC made only a subtle decrement ($<1\%$) in the performance or even made it better, indicating that these three features barely explained visual clustering alone. 
However, we cannot 
conclude that the bottom-three features (DC, DD, AC) have less significance than the top-three features (DSR, ED, SD) due to the existence of interplay between features (see \autoref{tab:ablation} (b)). For example, degradation resulting from switching off AC and DC together was substantially higher than the sum of degradation caused by switching off the features individually and was also higher than the largest degradation possible by removing a single feature (DSR). The same phenomenon occurs for ED and DC or DC and DSR. The interplay between DC and DSR was especially large compared to other combinations, validating the influence of the proximity factor in visual clustering. 
The significance of feature pairs suggests that future studies on visual clustering factors should focus on examining feature interplay in detail rather than analyzing features individually.

\begin{figure*}
    \centering
    \includegraphics[width=\linewidth]{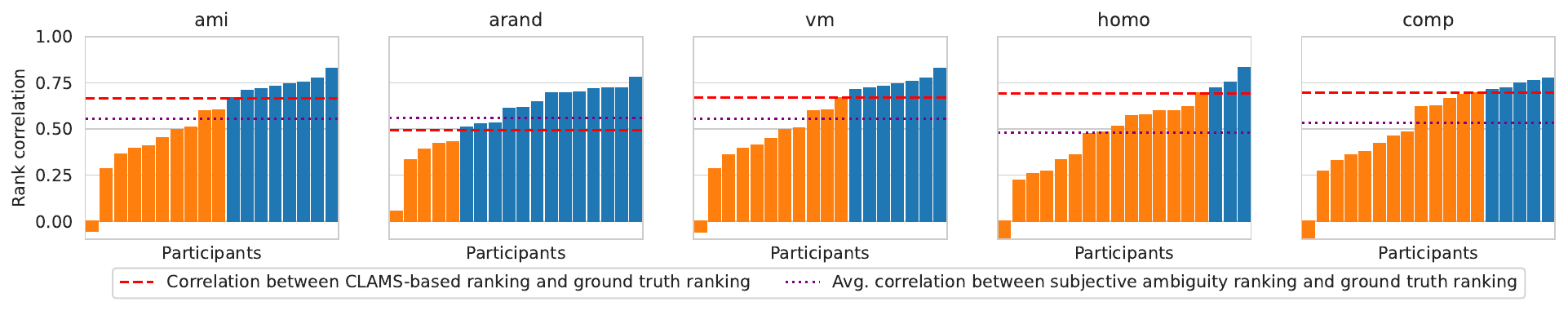}
    \vspace{-5mm}
    \caption{The comparison of the performance of \measure and human annotators (i.e., participants in our user study; \autoref{sec:mainstudydesc}) in estimating ground truth cluster ambiguity ranking.
             Each bar represents the rank correlation between the cluster ambiguity ranking made by the subjective response of a single participant and the ground truth ambiguity ranking.
             The orange and blue colors denote the participants who showed less and more accurate performance compared to \measure (red dashed line), respectively. 
             The purple dotted line depicts the average performance (i.e., rank correlation with ground truth ambiguity) of participants. 
             We found that \measure's performance is better than the average performance of participants but fails to outperform the performance made by the best annotator.\vspace{-4mm}
             }
    \label{fig:ambcoorbar}
\end{figure*}

\begin{figure}
    \centering
    \includegraphics[width=\linewidth]{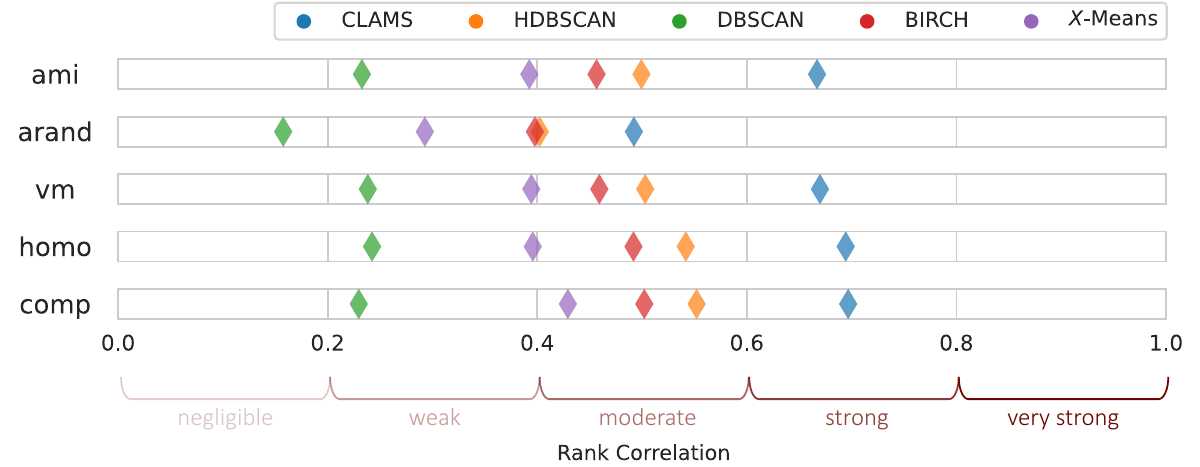}
    \caption{The rank correlation between ground truth ambiguity extracted based on five EVMs (\autoref{sec:groundtruthextract}) and the rankings made by computational ways to measure cluster ambiguity (\measure, variability of clustering techniques). Overall, \measure outperformed the competitors regardless of the used EVM, objectively having a strong correlation ($\rho > 0.6$) with ground truth~\cite{prion14csn} for the majority of the cases.  }
    \label{fig:ambrankcoor}
\end{figure}

\subsection{Accuracy Evaluation of \measure}

\label{sec:mainstudy}

We evaluated the accuracy of \measure by (1) constructing the ground truth cluster ambiguity ranking of scatterplots and (2) checking how well the ranking made by our measure matched the ground truth. 
The ground truth ambiguity of scatterplots was estimated based on a user study with 18 participants. 
Our results showed that \measure precisely estimated ground truth cluster ambiguity, outperforming the tested alternatives (e.g., human annotations, variability of clustering techniques).

\subsubsection{User Study for Constructing Ground Truth Ambiguity}

\label{sec:mainstudydesc}

\noindent
\textbf{Objectives and Tasks:}
Our user study aimed to construct the ground truth cluster ambiguity of scatterplots. We formulated two experimental tasks relevant to our research questions.

\begin{itemize}[noitemsep] 
    \item[\textbf{(T1)}] Lasso the clusters in the given scatterplot using the mouse.
    \item[\textbf{(T2)}] Subjectively determine the ambiguity of the scatterplot.
\end{itemize}
T1 aims to directly collect the visual clustering results of participants so that we could 
compute ground truth cluster ambiguity. We additionally asked participants to conduct T2 to check how well the subjective ambiguity annotated by individuals matches the ground truth ambiguity compared to \measure.

\noindent
\textbf{Procedure:}
The entire study consisted of three phases. 
We first collected the participants' non-identifying demographics.
We then asked the participants to conduct the two tasks listed above, viewing 60 selected stimuli (i.e., scatterplots) in random order.
Finally, participants answered the post-study interview questions: (1) \textit{``Which characteristics of the scatterplots do you think are most important in determining separation?''} (2) \textit{``What makes scatterplots ambiguous or clear based upon your response on the shown scatterplots?''}

\noindent
\textbf{Datasets and Preprocessing:} Our primary consideration in generating scatterplot stimuli was maximizing the diversity of cluster patterns (related to C2 in \autoref{sec:decon}). For this purpose, we first generated a large number of scatterplots with a high diversity of patterns. This was done by applying eight dimensionality reduction (DR) techniques ($t$-SNE~\cite{maaten08jmlr}, UMAP~\cite{mcinnes2020arxiv}, Densmap~\cite{narayan2021assessing}, Isomap~\cite{tenenbaum00aaas}, LLE~\cite{roweis00science}, MDS~\cite{kruskal64psycho}, PCA~\cite{pearson01pmjs}, and Random Projection) to 96 high-dimensional datasets~\cite{jeon22arxiv}. Except for PCA and MDS, we generated 20 scatterplots while randomly adjusting the hyperparameters to further diversify the patterns. We obtained $(20 \cdot 6 + 2) \cdot 96  = 11,712$ scatterplots in total.

We consecutively applied a stratified sampling, following Pandey et al.~\cite{pandey2016towards} and Abbas et al.~\cite{abbas19cgf}, to 
remove scatterplots with similar patterns~\cite{abbas19cgf}.  
To do so, we first grouped candidate scatterplots based on their number of clusters. We set the number as the optimal number of Gaussian components computed by GMM. We then computed Scagnostics~\cite{dang2014transforming} of scatterplots, representing each scatterplot as a vector consisting of Scagnostics scores (i.e., a vector abstracting the pattern of a scatterplot). Finally, we applied $K$-Means~\cite{likas03pr} to vectors representing each group of scatterplots and sampled scatterplots that were closest to centroids of resulting clusters. We 
set $K$ as 12. If the number of scatterplots in a group was less than 12, we sampled all scatterplots from the group.  
\rev{$K$ value is set to limit the number of sampled scatterplots to be around 100, making it possible to manually investigate each by eye.}
114 scatterplots were sampled. 
We manually sampled these scatterplots to minimize similar patterns, resulting in the final 60 scatterplots.

\noindent
\textbf{Participants:} 
\measure is designed for use in data analytics. We, thus, required participants to have experience in data analysis using scatterplots to ensure that they understood the concept of visual clustering. Based on snowball sampling \cite{goodman61ams}, we recruited 18 participants from the data visualization, human-computer interaction, and data mining communities (13 males and five females, aged 23-33 [27.2 $\pm$ 2.7]).
While 15 participants were graduate students, the remaining three were working professionals.
\rev{
Two participants self-reported as novice data analysts, two as intermediate, and 15 as experts. For familiarity with data analysis with scatterplots, 14 participants reported that they were at the expert level, and two participants said they were at an intermediate level. The remaining two participants reported themselves as novices.} 
We compensated each participant with the equivalent of \$20 for their time.

\noindent
\textbf{Apparatus:}
The experiment was conducted over a recorded Zoom session.
We developed a website for the study participants.
Participants were asked to access the website and share their screens so that the instructor could monitor and guide the experiment. 
\rev{As with the factor exploration study (\autoref{sec:preexp}), we fixed the stimuli size to 700px $\times$ 700px and constrained participants to use a laptop or desktop monitor.}
The studies were 45-60 minutes long. We instructed the participants on tutorials and experimental tasks, and conducted interviews at the end.

\begin{figure}[t]
    \centering
    \includegraphics[width=\linewidth]{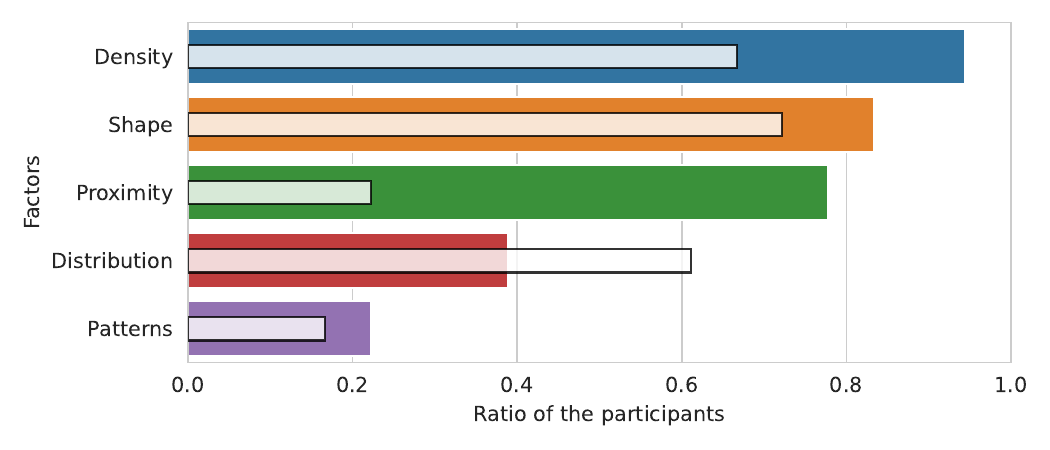} \vspace{-6mm}
    \caption{We identified the factors that participants considered when lassoing clusters and rating the scatterplot on a scale of very ambiguous to very clear in our main study (\autoref{sec:mainstudy}). 
    The length of the filled bars and white bars depicts the ratio of the participants who denote that the corresponding factor affects cluster separability and ambiguity, respectively.
    As the past studies~\cite{sedlmair2012taxonomy, sadahiro1997cluster} have demonstrated, we observed density, proximity, and shape as main factors.}
    \label{fig:qual_study}
\end{figure}

\subsubsection{Extracting Ground Truth Cluster Ambiguity} 

\label{sec:groundtruthextract}

We extracted ground truth cluster ambiguity for the sampled scatterplots by measuring the extent to which visual clustering results (i.e., lassoing result) differed by participant.
To compute the difference, we used external clustering validation measures (EVMs)~\cite{wu09kdd}, following previous research on visual clustering~\cite{desjardins07iui, jeon22arxivdistortion}. As EVMs can be applied only to a pair of clusters, we computed EVM scores of visual clustering results for each cluster pair and averaged the scores to get the final ground truth ambiguity. For EVMs, we used adjusted mutual information (\texttt{ami})~\cite{vinh10jmlr}, adjusted rand index (\texttt{arand})~\cite{steinley04pm}, v-measure (\texttt{vm})~\cite{rosenberg07emnlp}, homogeneity (\texttt{homo})~\cite{rosenberg07emnlp}, and completeness (\texttt{comp})~\cite{rosenberg07emnlp}.
\rev{We chose these measures as they are widely adopted in clustering and data mining communities \cite{rezaei16tkde, jeon22arxiv}.}
As a result, we obtained five ground truth cluster ambiguity rankings of scatterplots, where each corresponds to an individual EVM.
We only used EVMs that are ``adjusted'' so that the scores can be compared across different datasets~\cite{wu09kdd}.

\subsubsection{Results and Discussions} 

\label{sec:mainstudyresults}

\noindent
\textbf{Performance Analysis:}
The results validate the \measure's preciseness in estimating cluster ambiguity.
As seen in \autoref{fig:ambrankcoor}, the rankings made by \measure generally showed a strong correlation with the ground truth ranking ($\rho > 0.6$) based on a criterion of Prion and Haerling~\cite{prion14csn}. 
The correlation was, however, low for \texttt{arand}. 
This is because, unlike other EVMs that interpret clustering assignments as a probabilistic event, \texttt{arand} discretely ``counts'' the disagreement of clustering results and thus generates results that do not align with our probabilistic approach.
\measure also substantially outperformed clustering techniques in terms of predicting ground truth for all EVMs, verifying that \measure is the best of the tested computational options. 

As depicted in \autoref{fig:ambcoorbar}, 
\measure showed better performance in estimating ground truth cluster ambiguity compared to the average human annotators. 
Except for the case of \texttt{arand}, \measure always showed a better correlation with ground truth (red dashed line) compared to the average correlation of human annotators (purple dotted line) regardless of the EVM choice.
On average, 9.8 out of 18 participants (54\%; the proportion of orange bars) made less accurate predictions compared to \measure.
This result indicates that it is difficult to estimate ambiguity for the judgment of a single person. Therefore, our automatic solution offers increased value in accurately evaluating ambiguity.

\noindent
\textbf{Post-Study Interview:}
As described in \autoref{sec:mainstudydesc}, we interviewed participants 
to identify the reasoning behind their responses. We found that the main factors behind the lassoing (i.e., visual clustering) and determining the scatterplot ambiguity levels were density, proximity, shape, and distribution (as shown in \autoref{fig:qual_study}). We elaborate on these findings in detail in \autoref{sec:discuss}.

%% file: sections/05_applications.tex
\section{Applications}
\label{sec:appl}

We introduce two applications of \measure that validate the effectiveness, generalizability, and applicability of \measure.
We present that 
\measure can further optimize nonlinear dimensionality reduction embeddings to reduce cluster ambiguity while maintaining accuracy (\autoref{sec:ambreduce}). 
We then demonstrate that the measure can be used to find reliable datasets for benchmarking clustering techniques (\autoref{sec:reliable}).

\subsection{Reducing Cluster Ambiguity of Nonlinear Dimensionality Reduction Embeddings}

\label{sec:ambreduce}

\subsubsection{Problem Statement and System Design}

A common way to analyze the cluster structure of high-dimensional data is to use dimensionality reduction (DR)~\cite{nonato19tvcg} techniques, e.g., $t$-SNE~\cite{maaten08jmlr}, Isomap~\cite{tenenbaum00aaas}, or UMAP~\cite{mcinnes2020arxiv}, which synthesizes 2D representation (i.e., embedding) that preserves the original characteristics of the input high-dimensional data. 
The analysis is usually done by depicting the embedding as a scatterplot and conducting visual clustering. 
Recently, nonlinear DR techniques~\cite{lee07springer} that can reveal the complex data manifold have been widely developed~\cite{jeon22vis, fu19kdd,maaten08jmlr, mcinnes2020arxiv} and used for cluster analysis~\cite{becht19nature, fujiwara22arxiv}.  

Embedding varies by choice of DR technique or hyperparameter, where each can lead to significantly different analysis results. A typical way to address this problem is to optimize embeddings by assessing their accuracy in representing the original data. 
Diverse metrics to measure the accuracy of DR embeddings~\cite{jeon22tvcg, lee07springer, nonato19tvcg} have been developed for the purpose. 
However, an optimized embedding can still have an ambiguous cluster structure, which
can potentially harm the reliability of the analysis despite high accuracy (\autoref{sec:intro}). 
For the valid analysis of high-dimensional data, an embedding having both high accuracy and low cluster ambiguity is required.  

Here, we introduce an optimization system that can discover DR embeddings satisfying both high accuracy and low ambiguity, which we call \ambreducer. Given the input high-dimensional data $Z$, DR technique $t$, and DR accuracy metric $m$, \ambreducer first finds the \textit{intermediate} embedding that maximizes accuracy.
This is done by searching a hyperparameter setting $h_1 = \arg\max_{h\in H}{m(t(Z, h))}$, where $H$ denotes the set of every possible hyperparameter setting and $t(Z, h)$ represents the embedding made with $t$ using hyperparameter setting $h$. We use Bayesian optimization~\cite{snoek12nips} over $H$ while maximizing $m(t(Z, h))$. Then, the system searches for the \textit{final} embedding that maintains accuracy while reducing cluster ambiguity compared to the intermediate embedding. 
Formally, this is done by finding hyperparameter setting $h_2 = \arg\max_{h\in H}{loss(h, h_1)}$, where $loss$ is defined as
\resizebox{\linewidth}{!}{%
\begin{minipage}[t]{1.15\linewidth}
\vspace{-3pt}
\[
loss(h, h_1) = \left\{ \begin{array}{ccc}
   CA(t(X, h))    & \mbox{ if } &\mid m(t(Z, h_1)) - m(t(Z, h)) \mid < \tau \\
   \infty       & \mbox{ if } &\mid m(t(Z, h_1)) - m(t(Z, h)) \mid > \tau 
\end{array}  \right..
\]
\vspace{6pt}
\end{minipage}}
Here, $CA$ denotes \measure score, and $\tau$ denotes the threshold parameter, which can be set by users. $\tau$ adjusts the tradeoff between accuracy and cluster ambiguity. If we set small $\tau$, the final embedding made by $h_2$ tends to have high accuracy but has fewer chances to reduce ambiguity. 
\rev{The proper value of $\tau$ highly depends on the target application and domain. For example, the more analysts involved, the greater $\tau$ should be to give more room to reduce ambiguity. Automatically determining $\tau$ is important future work that extends the applicability of \ambreducer.}

\subsubsection{Evaluation}

\label{sec:ar_eval}

\begin{figure}[t]
    \centering
    \includegraphics[width=\linewidth]{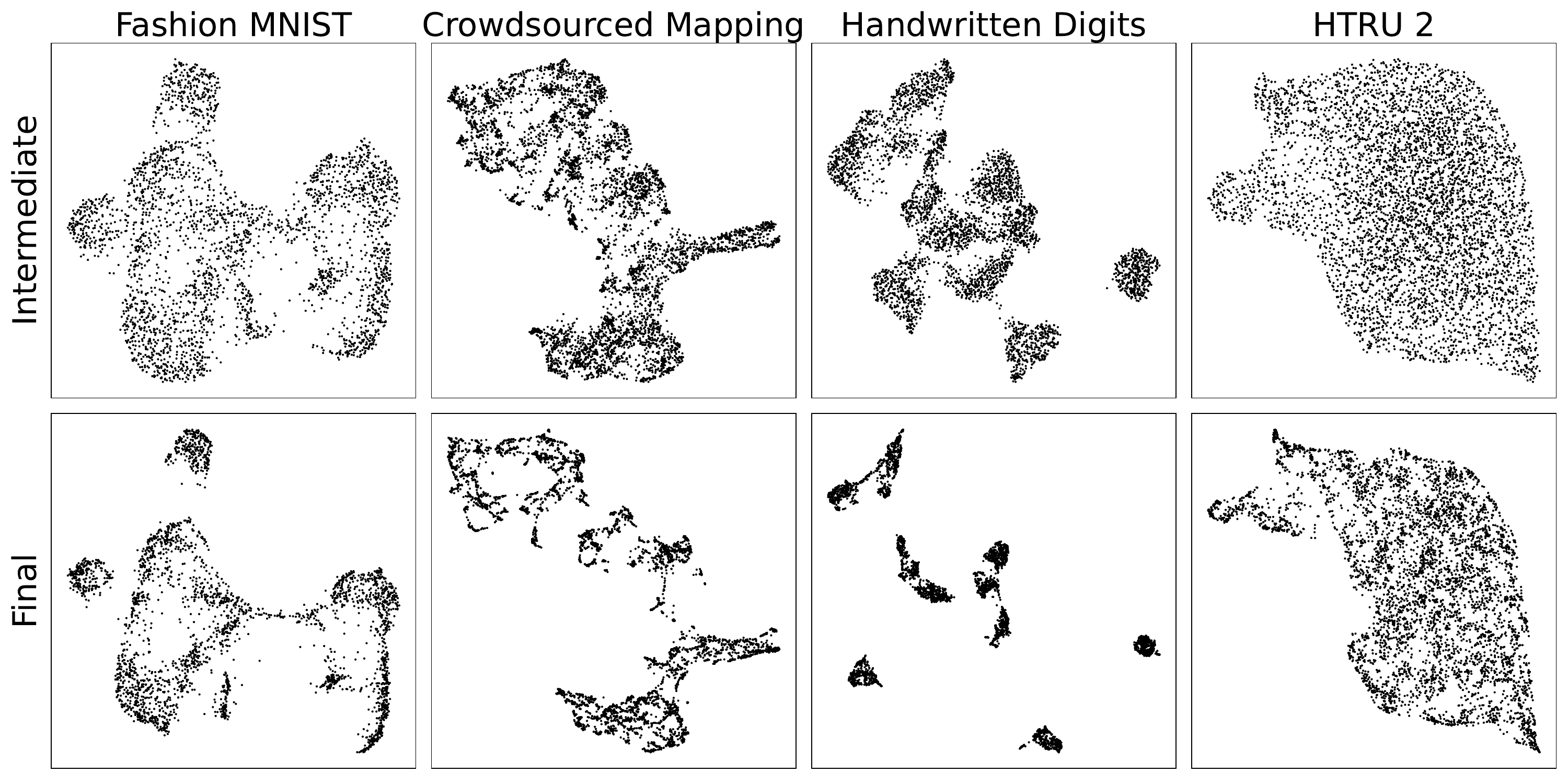} \vspace{-5mm}
    \caption{Dimensionality reduction embeddings optimized solely based on the accuracy (top row; intermediate results of \ambreducer) and those optimized considering both accuracy and cluster ambiguity (bottom row; final results of \ambreducer). Our experiment shows that the accuracy of the final results has no significant difference compared to the intermediate results in terms of accuracy but has a substantially smaller amount of cluster ambiguity (\autoref{sec:ar_eval}, \autoref{fig:ar_exp}). \vspace{-4mm}}
    \label{fig:ar_scatter}
\end{figure}

\begin{figure}[t]
    \centering
    \includegraphics[width=\linewidth]{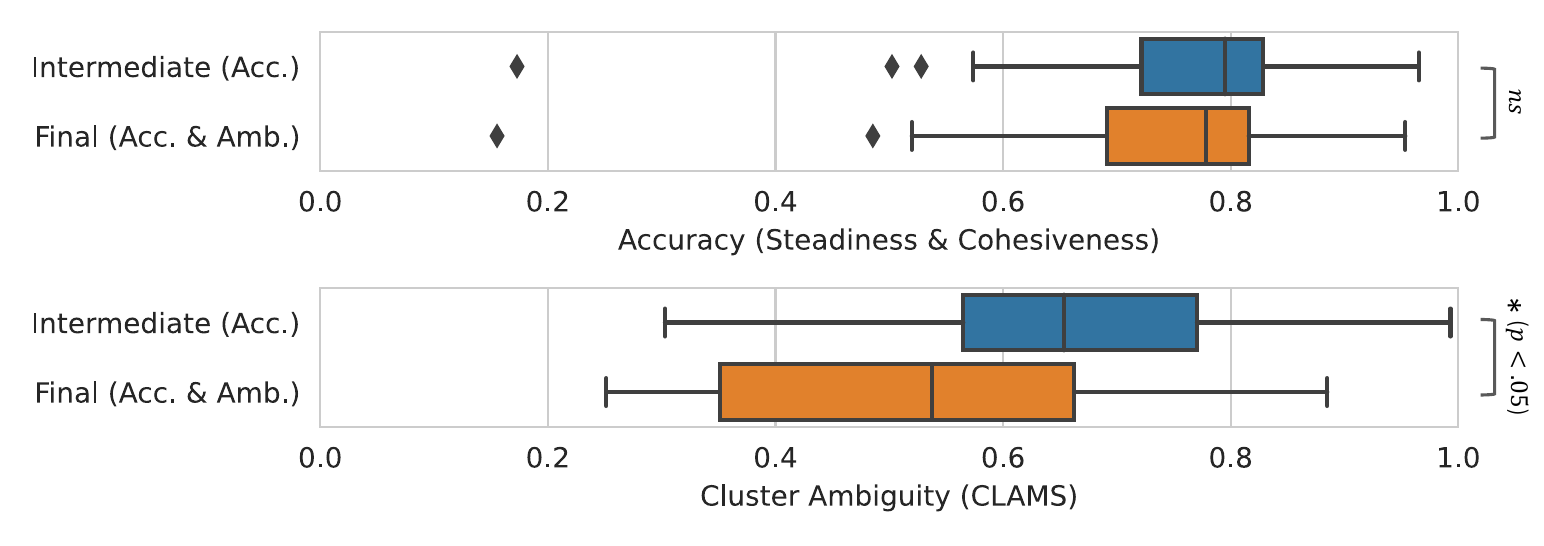} \vspace{-7mm}
    \caption{The results of our evaluation on \ambreducer (\autoref{sec:ar_eval}). 
    We compared the accuracy and cluster ambiguity of the intermediate (optimized solely on accuracy; blue) and final (optimized based on both accuracy and cluster ambiguity; orange) embeddings made by \ambreducer.
    We found that there is no significant difference (ns) between the final and intermediate results in terms of accuracy (top), while the cluster ambiguity is significantly reduced ($p<.05$) in the final results (bottom). \vspace{-5mm}}
    \label{fig:ar_exp}
\end{figure}

\begin{figure*}
    \centering
    \includegraphics[width=\linewidth]{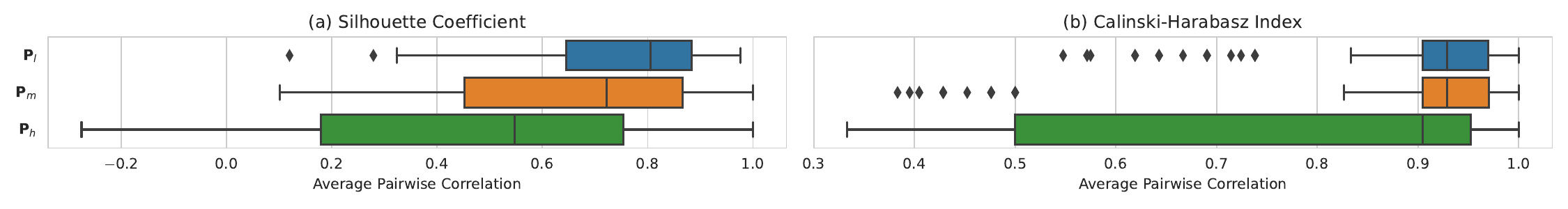}
    \vspace{-6.5mm}
    \caption{
    Results of the experiment demonstrating \measure's effectiveness in picking reliable clustering benchmark datasets (\autoref{sec:reliable}). 
    We first prepared a set of datasets (bottom-20 ($\mathbf{P}_l$; blue), middle-20 ($\mathbf{P}_m$; orange) or top-20 ($\mathbf{P}_h$; green) subset of 60 scatterplots we used in the main study based on \measure score) and clustering metrics (Silhouette Coefficient (a) or Calinski-Harabasz Index (b)).
    For each set, we compute the rankings of clustering techniques over individual datasets within the set based on clustering metrics.
    We then compute the within-set stability by assessing rank correlations of the rankings pairwisely.
    In general, the set consisting of datasets with less ambiguity showed better stability, confirming the hypothesis that datasets with less cluster ambiguity are more reliable for benchmarking clustering techniques.
    \vspace{-4.5mm}
    }
    \label{fig:app_clustering}
\end{figure*}

\noindent 
\textbf{Objectives and Design:}
We wanted to verify the effectiveness of \ambreducer.
We hypothesized that \ambreducer substantially reduces cluster ambiguity while maintaining accuracy.
To validate the hypothesis, we prepared 30 high-dimensional datasets, sampled from the datasets we used in our main study \autoref{sec:mainstudy}, and applied \ambreducer.
We collected the intermediate (based on $h_1$) and final (based on $h_2$) embeddings, and checked how the accuracy and cluster ambiguity varied between the two groups.
\autoref{fig:ar_scatter} depicts the subset of two groups.

\noindent
\textbf{Hyperparameter Setting:}
Our hyperparameter setting was as follows. 
For the DR technique, we used UMAP, as UMAP is widely used for cluster analysis in practice~\cite{becht19nature, sanchez2020dimensionality}.
We used Steadiness \& Cohesiveness~\cite{jeon22tvcg} (S\&C) as accuracy measures, as they are specially developed to measure the accuracy in preserving cluster structure.
We used the F1 score of S\&C to represent the accuracy of DR, which ranges from 0~to~1.
Finally, we arbitrarily set $\tau$ as 0.05. 

\noindent
\textbf{Results and Discussions:}
We used a $t$-test to check whether there is a significant accuracy (measured by S\&C) or cluster ambiguity (measured by \measure) difference between the final and intermediate embeddings.
Although there was no significant difference  in terms of accuracy ($t(58)=.455, p = .650$), we found that the cluster ambiguity of the final embeddings was significantly lower than the one of the intermediates ($t(58)=2.55, p < .05$). 
The result verifies our hypothesis, verifying the effectiveness of \ambreducer and the applicability of \measure.

\subsection{Finding Reliable Datasets for Clustering Benchmark}
\label{sec:reliable}
We demonstrate an experiment validating \measure's effectiveness in selecting reliable datasets for benchmarking clustering techniques~\cite{liu10icdm}.
We hypothesized that if the cluster ambiguity of a dataset (a scatterplot in this case) is high, then the clustering benchmark using the dataset is unreliable. 
Our assumption is that if the dataset is ambiguous, which means that there is no explicit cluster structure to be found, the performance of clustering techniques to ``find the cluster'' will be insufficiently compared. 
Thus, clustering benchmarking with the ambiguous dataset will likely provide an arbitrary ranking of techniques, which harms the credibility of the evaluation. 
To verify the hypothesis, we prepared three sets of datasets having different cluster ambiguity levels and checked the stability of the rankings made by different datasets in each set. 
The detailed explanation of the experiment is as follows:

\noindent
\textbf{Objectives and Design:}
We wanted to check whether the cluster ambiguity of benchmark datasets affects the stability of the ranking of clustering techniques, which we use as a proxy for the reliability of the evaluation. 
We prepared three sets of datasets with the same cardinality: $\mathbf{P}_h$, $\mathbf{P}_m$, and $\mathbf{P}_l$, with each having high, middle, and low cluster ambiguity, respectively. 
For each set, we obtained rankings of clustering techniques, where each was made based on an individual dataset. We then checked how the rankings correlate with each other by averaging the pairwise rank correlation computed by Spearman's $\rho$.

\noindent
\textbf{Datasets:}
We used 60 scatterplots we gathered in the main evaluation for \measure (see \autoref{sec:mainstudy}). We sorted the scatterplots based on \measure score in descending order, and then assigned top-20, middle-20, and bottom-20 scatterplots to $\mathbf{P}_h$, $\mathbf{P}_m$, and $\mathbf{P}_l$, respectively.  

\noindent
\textbf{Clustering Techniques:}
We used eight techniques: HDBSCAN~\cite{campello13kdd}, DBSCAN, $K$-Means, $X$-Means~\cite{pelleg00icml}, Birch~\cite{zhang96sigmod}, and Agglomerative clustering~\cite{mullner11arxiv} with single, average, and complete linkage.

\noindent
\textbf{Measurements:}
We used the Silhouette coefficient~\cite{rousseeuw87jcam} and Calinski-Harabasz index~\cite{calinski74cis} to measure the performance of clustering techniques. To ensure that the evaluation reflected the maximum capability of clustering techniques, we ran Bayesian optimization and used the best score obtained while following the hyperparameter range setting of Jeon et al.~\cite{jeon22arxiv}.
Refer to Appendix C for detailed settings.

\noindent
\textbf{Results and Discussions:}
\autoref{fig:app_clustering} depicts the results.
We used a one-way ANOVA to analyze how the rank correlation scores vary due to the used set of datasets ($\mathbf{P}_l$, $\mathbf{P}_m$, $\mathbf{P}_h$). We used Tukey's HSD for post hoc analysis. 
For both the clustering metrics, the set of datasets had a significant effect (Silhouette: $F(3,567)=54.1, p < .001$; Calinski-Harabasz: $F(3, 567)=44.0, p < .001$). Post-hoc analysis revealed that the rank correlation scores over $\mathbf{P}_l$ and $\mathbf{P}_m$ were significantly higher than the ones over $\mathbf{P}_h$ for both metrics ($p < .001$ for every case). Meanwhile, scores over $\mathbf{P}_l$ were significantly higher than the ones over $\mathbf{P}_m$ for the Silhouette coefficient ($p < .001$), but not for the Calinski-Harabasz index ($p=0.301$). Such results indicate that using datasets with less cluster ambiguity makes the clustering evaluation more stable, confirming our hypothesis and validating the effectiveness of \measure in picking reliable datasets for the clustering benchmarks.


%% file: sections/06_conclusion.tex
\section{Discussion}
\label{sec:discuss}

Our approach presents a system to identify and assess the ambiguity of cluster structure in scatterplots. We provided preliminary steps toward a comprehensive automation system 
to simulate human judgments 
in scatterplots. In this section, we discuss the implications of our measure design and experiments 
including opportunities for future work.

%
%

\subsection{Proxy for Human Perception}
In this work, we provided preliminary steps toward developing the model to identify and rank intrinsic variability in visual perception. 
Evaluating variability based on human participation is not always feasible, scalable, or robust considering 
the diversity of data characteristics, visualization designs, and other factors that might influence people's perceptions. There are simply too many sources of variance to account for.
\measure, which automatically generates a score simulating human judgments, offers a scalable and robust alternative to an approach that attempts to account for every possible source of variance. 
%
Our work demonstrates the promise of using statistical modeling to infer patterns over a corpus of human estimates.  
%
Extending our 
approach to other perceptual tasks (e.g., outlier detection)
would 
bring the notion of perceptual variability to more complex and practical applications \cite{moritz2018formalizing}.


Toward this end, we provide a \href{http://www.clusterambiguity.dev.s3-website.ap-northeast-2.amazonaws.com/}{demo interface} that measures and explains cluster ambiguity. 
We believe that the interface will enhance the usability of \measure and moreover serve as a preliminary step toward future applications considering perceptual variability.

\subsection{Credibility of the User Study-Based Design Process}

\label{sec:credibility}

While designing \measure, we identified several factors that play a vital role in cluster perception through a user study (\autoref{sec:preexp}), and built a regression module estimating human cluster perception based on study findings (\autoref{tab:feature_eng}). These factors support building a robust module that accurately reflects how people identify clusters in data analysis (\autoref{sec:regmodel}).
Our ablation study (\autoref{sec:regmodeleval}) verifies the importance of these factors, validating the credibility of using human strategies to inform model parameters. 
Moreover, the main study (\autoref{sec:mainstudy}) showed that the measure built upon such a design process reliably estimates the ambiguity of scatterplots with a wide range of cluster patterns.

Our study asked participants about how the characteristics of datasets influence the ambiguity of a scatterplot (see \autoref{sec:mainstudy}). The findings from the interview validate the reliability of our design process. As shown in \autoref{fig:qual_study}, participants identified the factors that we have considered in our model (see \autoref{tab:feature_eng}), such as density, proximity, shape, and how clusters are located and related to each other (i.e., distribution). 
Note that such results also match well with prior findings~\cite{sedlmair2015data, sadahiro1997cluster,quadri21tvcg}. 

\subsection{\textit{Ambiguity} of Ambiguity}

While our interview revealed common factors that influence cluster perception (\autoref{sec:credibility}), it also showed that the perceived influence of different factors on ambiguity varies among participants. In other words, the definition of cluster ambiguity is \textit{ambiguous} to each participant. 
This observation depends on what factors are given more importance by different participants when determining a scatterplot's ambiguity.
For example, P01 said, \textit{"I find the shape of the group of points to be the main reason in identifying and separating clustering. If they have salient separable shapes, they are more clear"}, and P07 noted, \textit{"Proximity and concentration of the points help me separate the clusters quickly"}. 
\autoref{fig:qual_study} also supports this finding, showing that no single factor received complete agreement from the participants (see the white transparent bars).
Considering the variability across participants,
we conclude that ambiguity cannot be determined by a single person but must instead reflect a population of individuals. 
The fact that the performance of human annotators (i.e., participants) in predicting ground truth ambiguity varied in our main study (\autoref{sec:mainstudy}) supports this claim.
This observation indicates that, inevitably, multiple subjects are required for assessing ambiguity through human resources, thereby underscoring the importance of our automated solution.

\subsection{Limitations and Future Work}
\measure outperformed computational approaches and more than 50\% of participants (\autoref{sec:mainstudy}) in precisely estimating cluster ambiguity. However, there 
are several opportunities to improve \measure.
For example, we found that \measure erroneously considers scatterplots with more numbers to have less ambiguity, as seen in \autoref{fig:topbottom}---the top eight clear scatterplots (top row) generally have more clusters than the top eight ambiguous ones (bottom row). We quantitatively demonstrate this bias in Appendix D.
\rev{We can also improve the method of aggregating pairwise ambiguity scores, which is currently a na{\"i}ve average. Considering cluster topology or their pairwise distances during the aggregation may better reflect human visual perception.}

\rev{
Another open direction is to generalize \measure. We can extend \measure to consider visual encoding (e.g., size, shape, and color of marks) of scatterplots by revising the features feed into our regression module. We may also improve our measure to deal with nested clusters by adopting hierarchical clustering algorithms \cite{mullner11arxiv} in place of GMM. We are also interested in
}
broadening the concept of ambiguity to encompass general visualization. 
For example, we can model the perceptual variability that may arise when examining cluster structures of high-dimensional data via scatterplot matrices or parallel coordinates. Moreover, we aim to develop a model estimating ambiguity in various perceptual tasks, including outlier detection and trend analysis.

\section{Conclusion}

We introduce \measure, a VQM for estimating the cluster ambiguity of a monochrome scatterplot, which originally required expensive human resources to compute. 
To serve as a proxy for human perception across a wide range of cluster patterns, our measure is designed based on a qualitative user study and is trained over perceptual data.
Through quantitative evaluations, we verified that \measure outperforms automatic competitors while showing competitive performance with human annotators. Our research findings not only demonstrate current applications but also invite discourse on potential future applications that capitalize on the concept of ambiguity.
In summary, our work represents a \rev{significant} advancement toward developing a comprehensive framework that elucidates the phenomenon of ambiguity in visualization. 

\newpage